\documentstyle[psfig]{mn}

\newcommand\mzon   {M$_{\odot}$}
\newcommand\pp     {$\pm$}
\newcommand\pers     {s$^{-1}$}
\newcommand\micros  {$\mu$s}
\newcommand\Lunit   {erg s$^{-1}$}
\newcommand\funit   {erg s$^{-1}$ cm$^{-2}$}

\def\degr{\hbox{$^\circ$}}

\title[RXTE observations of 4U 1957+11]{4U 1957+11: a persistent
low-mass X-ray binary and black-hole candidate in the high state?}

\author[Wijnands et al.]  {Rudy Wijnands$^1$\thanks{Chandra Fellow;
     Email:rudy@space.mit.edu}, Jon M. Miller$^1$, Michiel van der
     Klis$^2$\\ 
     $^1$ Center for Space Research, MIT, NE80-6055, 77 Massachusetts Avenue,
     Cambridge, MA 02139-4307, USA\\
     $^2$ Astronomical Institute ``Anton Pannekoek'', University of
     Amsterdam, Kruislaan
     403, NL-1098 SJ Amsterdam, The Netherlands}

\begin{document}
\maketitle

\begin{abstract}
We report on several pointed {\it Rossi X-ray Timing Explorer}
observations of the enigmatic low-mass X-ray binary (LMXB) 4U 1957+11
at different X-ray luminosities. The luminosity of the source varied
by more than a factor of four on time scales of months to years. The
spectrum of the source tends to get harder when its luminosity
increases. Only very weak (1\%--2\% rms amplitude; 0.001--10 Hz; 2--60
keV) rapid X-ray variability was observed during the observations. A
comparison of the spectral and temporal behaviour of 4U1957+11 with
other X-ray binary systems, in particular LMC X-3, indicates that 4U
1957+11 is likely to be a persistent LMXB harboring a black hole and
which is persistently in the black-hole high state. If confirmed, it
would be the only such system known.

\end{abstract}


\section{Introduction \label{intro}}

Low-mass X-ray binaries (LMXBs) are systems in which a compact object
(either a neutron star or a black hole) accretes matter from a
low-mass ($<$1 \mzon) companion star. Approximately 150 LMXBs are
known (van Paradijs 1995; Liu, van Paradijs, van den Heuvel 2001) and
for many of these systems the nature of the compact object could be
uniquely identified as a neutron star based on the observation of
thermonuclear flashes from the star's surface (so-called type-I X-ray
bursts) or from X-ray pulsations. For several LMXBs (all X-ray
transients) the nature of the compact object was deduced to be a
black hole based on the large mass-function (and consequently the
large lower limit on the mass of the compact object) obtained from
those systems when they were in their quiescent state (see, e.g.,
Filippenko et al. 1999 and McClintock et al. 2001 and references
therein). For the remaining LMXBs the nature of the compact is still
unknown. However, the possible nature is often deduced from
similarities of those systems with LMXBs for which the nature of the
primary could be determined.  But neutron star and black-hole LMXBs
are very similar with respect to their X-ray spectral (e.g., Barret \&
Vedrenne 1994; Barret et al. 1996) and X-ray timing properties (e.g.,
van der Klis 1994a, 1994b; Wijnands \& van der Klis 1999) and
classifications based on resemblances are difficult and subject to
errors.

One of the persistent LMXBs for which the exact nature of the compact
object is still a subject of debate is 4U 1957+11.  Its soft X-ray
spectrum led White \& Marshall (1984) to suggest that 4U 1957+11 is a
candidate to harbor a black hole (a black-hole candidate or
BHC). However, such a soft spectrum is not a strong conclusive
argument and can be interpreted in different ways. For example, the
spectral studies reported in the literature both give evidence for a
neutron star ({\it Ginga}: Yaqoob, Ebisawa, \& Mitsuda 1993; {\it
EXOSAT}: Singh, Apparao, \& Kraft 1994) or a black hole ({\it EXOSAT}:
Ricci, Israel, \& Stella 1995) primary. The study performed by Nowak
\& Wilms (1999; using data obtained with the X-ray satellites {\it
Rosat}, {\it ASCA}, and the {\it Rossi X-ray Timing Explorer} [{\it
RXTE}]) showed that the spectrum of 4U 1957+11 can be understood in
different ways and that the nature of the compact object remains
elusive.

The optical counterpart of 4U 1957+11 (V1408 Aquilae) was discovered
by Margon, Thorstensen, \& Bowyer (1978) and its orbital period of
9.33 hr by Thorstensen (1987).  Hakala, Muhli, \& Dubus (1999) studied
V1408 Aquilae in more detail and reported that the optical light curve
pulse shape over the orbital period changed between their study and
the study of Thorstensen (1987).  They proposed that these changes are
due to an evolving accretion disc structure.  From their models a
large inclination angle of 70\degr--75\degr~was obtained, which Nowak
\& Wilms (1999) combined with the 117 day period they reported in the
{\it RXTE} All Sky Monitor (ASM) light curve of 4U 1957+11 to propose
that the X-ray spectrum and its variations are due to a warped
precessing accretion disc.

In this paper, we present data of 4U 1957+11 obtained with the
proportional counter array (PCA) on board {\it RXTE} when the source
was at luminosities up to four times higher than during the
observations reported by Nowak \& Wilms (1999). We also analyzed a
more extensive {\it RXTE}/ASM data set and conclude the long-term
behaviour significantly changes with time.  We discuss our spectral and
timing results with respect to the nature of the compact object.

\section{Observations}

\begin{figure}
\begin{center}
\begin{tabular}{c}
\psfig{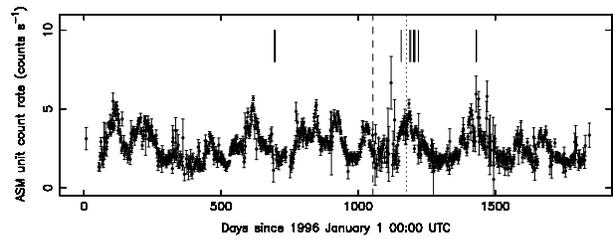}
\end{tabular}
\caption{The 1.5--12 keV {\it RXTE}/ASM count rate curve in 2 days
rebinning up to 18 January 2001.  The solid lines correspond to the
dates of the observations listed in Table \ref{tab:log} (the first
line corresponds to the AO2 data and the others to the AO4 data). The
dotted line marks the end of {\it RXTE}/PCA gain epoch 3 and the start
of epoch 4. The data up to the dashed line was already presented by
Nowak \& Wilms (1999) and used in their analyze.
\label{fig:lc_asm} }
\end{center}
\end{figure}

During Cycle 2 of {\it RXTE}, 4U 1957+11 was observed for $\sim$29
ksec when the source was at low luminosities (Fig.~\ref{fig:lc_asm};
see also Nowak \& Wilms 1999; hereafter referred to as the AO2
observations).  To study the spectral and timing behaviour of this
source at higher luminosity levels, we had proposed to perform Cycle 4
{\it RXTE} TOO observations on this source when its count rate would
exceed 3 ASM counts \pers.  This programme was approved, and
observations (hereafter referred to as the AO4 observations) were
indeed triggered, resulting in a total of 38.6 ksec of data (see Table
\ref{tab:log} for a log of the observations). We reanalyzed the AO2
data to obtain an homogeneous analysis for all observations.

During all observations data were collected in the Standard 1 (1/8 s
time resolution in one photon energy channel for the energy range
2--60 keV) and the Standard2f (129 channels for 2--60 keV; 16 s time
resolution) modes. Simultaneously, data were also collected during the
AO2 data with the GoodXenon1\_16s and GoodXenon2\_16s modes (together
they have 256 channels for 2--60 keV and a time resolution of $\sim$ 1
\micros). During the AO4 observations, data were collected in one
event mode (E\_125us\_64M\_0\_1s: 128 \micros~ time resolution in 64
channels from 2--60 keV), one single bit mode (SB\_125us\_0\_249\_1s:
128 \micros~ time resolution in 1 channel from 2--60 keV), and a burst
trigger (TLA\_1s\_10\_249\_1s\_10000\_F) and catcher
(CB\_8MS\_64M\_0\_249\_H) mode. The last three modes were active in
order to study type-I X-ray bursts (the single bit mode for high time
resolution and the catcher mode for high spectral resolution at modest
time resolution) in case 4U 1957+11 should harbor a neutron
star. However, no bursts were found and we do not discuss the data
obtained in these modes further in our paper.

The five PCA detectors (called proportional counter units or PCUs)
each have a slightly different energy response. Due to aging of the
electrodes and processes as gas leakage, the gain of the individual
detectors drifts slowly in time. More sudden changes in the gain are
caused by changes in the high voltage settings (which are occasionally
made in order to preserve the detectors), resulting in (so far) four
major gain epochs.  The AO2 observations and the first one of the AO4
observations (observation number 5 in Table~\ref{tab:log}) were taken
during gain epoch 3 and the other observations during gain epoch
4. Not all of the five detectors were always on during our
observations, resulting in a different number of active detectors
(between 2 and 5, see Table~\ref{tab:log}) throughout our study.

\section{Analysis and results}

\subsection{The {\itshape RXTE}/ASM light curve \label{section:asm}}

We used the quick-look one-day averaged {\it RXTE}/ASM
data\footnote{the quick-look data can be obtained from
http://xte.mit.edu/ASM\_lc.html and is provided by the ASM/{\it RXTE}
team. See Levine et al. (1996) for a detailed description of the ASM.}
of 4U 1957+11 in order to study its long-term variability.  The
resulting long-term ASM light curve of 4U 1957+11 is shown in
Figure~\ref{fig:lc_asm}.  The source smoothly varies between $\sim$1
and $\sim$6 ASM count \pers, with clear episodes of low and high
luminosities. In this figure, we have indicated when the pointed {\it
RXTE} AO2 (first solid vertical line) and our AO4 observations (the
other solid vertical lines) were performed. Clearly our strategy to
observe 4U 1957+11 at high luminosities succeeded, resulting in 38.6
ksec of data during which the luminosity of the source was 2--4 times
higher than during the AO2 observations.

\begin{figure}
\begin{center}
\begin{tabular}{c}
\psfig{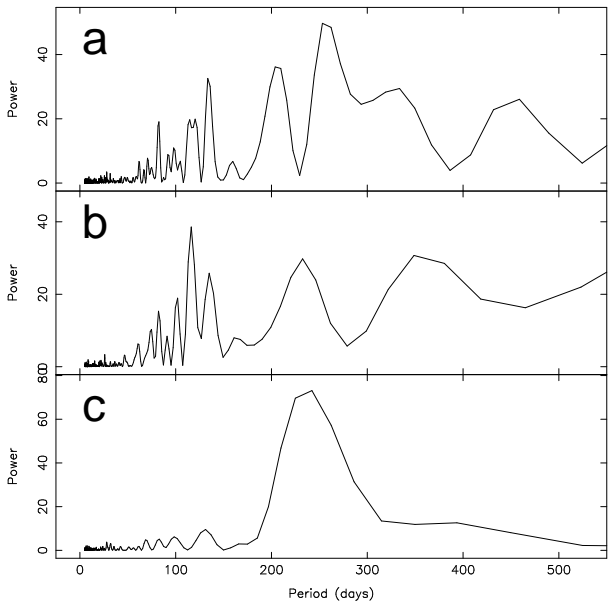}
\end{tabular}
\caption{Lomb-Scargle diagrams for 4U 1957+11 of ({\it a}) all {\it
RXTE}/ASM data available (up to 2001 January 18; using the light curve
which was also used to make Figure \ref{fig:lc_asm}), ({\it b}) the
data used by Nowak \& Wilms (1999; up to 1998 November 20), and ({\it
c}) the data obtained between 1998 November 20 and 2001 January
18.\label{fig:ls_diagram} }
\end{center}
\end{figure}

Nowak \& Wilms (1999) reported a 117 day period in the {\it RXTE}/ASM
long-term X-ray light curve of 4U 1957+11. They used the available
data up to 1998 November 20. Since then, a significantly larger data
set has become available and therefore we decided to analyze all ASM
data available up to the writing of this paper (2001 January 18).  We
made a Lomb-Scargle diagram (Lomb 1976; Scargle 1982) of the ASM data
(see Fig. ~\ref{fig:ls_diagram}{\it a}). This figure clearly shows
multiple peaks above 100 days. The most significant peak is the one at
250--260 days, but several other peaks between 100 and 400 days are
also present.  These peaks are in the range of the ones found by Nowak
\& Wilms, but they are not exactly similar. This could indicate that
no complete stable period is present in the system.

In order to study this possibility, we made Lomb-Scargle diagrams of
the data used by Nowak \& Wilms (1999; up to 20 November 1998;
Fig. \ref{fig:ls_diagram}{\it b}) and of the data taken after that (up
to 18 January 2001; Fig. \ref{fig:ls_diagram}{\it c}). The 117 day
peak is clearly the most significant in the first data set, however,
in the second data set this peak is nearly absent but a dominant peak
is present around 250 days. This result strongly suggest that the
long-term variability behaviour of 4U 1957+11 is more complicated than
suggested by the 117 day period reported by Nowak \& Wilms (1999).

\subsection{The {\itshape RXTE}/PCA light curve \label{section:lc_pca}}

\begin{figure}
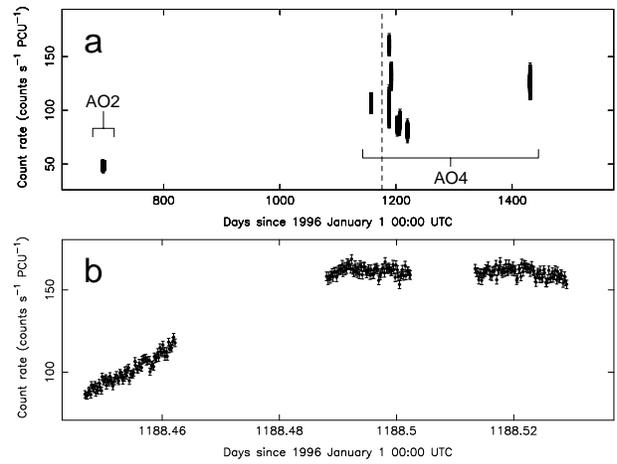

\begin{center}
\begin{tabular}{c}
\psfig{figure=f3a.eps,width=8cm,angle=-90}\\
\psfig{figure=f3b.eps,width=8cm,angle=-90}
\end{tabular}
\caption{The 2--60 keV {\it RXTE}/PCA count rate curves, with ({\it
a}) the total count rate curve and ({\it b}) a blow up of observation
40044-01-02-01 (observation 6) showing the rising phase and the
plateau. The count rates are background subtracted but not dead-time
corrected ($<$0.5\%). The dashed line in ({\it a}) indicates when the
high voltages of the PCUs were changed and marks the end of gain epoch
3 and the start of gain epoch 4.
\label{fig:lc_pca} }
\end{center}
\end{figure}

We used the Standard2f data in order to create a {\it RXTE}/PCA light
curve for the total {\it RXTE}/PCA energy range of 2--60 keV (see
Fig.~\ref{fig:lc_pca}{\it a}). The count rates were background
subtracted but no dead-time correction was applied. However, we
estimate that the correction factor would be $<$0.5\%, which is
considerable smaller than the errors on the count rates due to
counting statistics. From Figure~\ref{fig:lc_pca}{\it a} and from
Table~\ref{tab:log}, it can been seen that we have obtained {\it
RXTE}/PCA data of 4U 1957+11 at a count rate which is up to 4 times
higher than its count rate observed during the AO2 observations. This
allows us to study the 3--20 keV X-ray spectrum and the rapid X-ray
variability of this source at different luminosities.

The count rates during the individual observations vary only slightly
(typically between 5\%--15\%).  However, one observation (observation
6) behaved differently. In Figure~\ref{fig:lc_pca}{\it b}, the PCA
light curve of this observation is shown. At the beginning of the
observation, the source was at $\sim$85 counts \pers~PCU$^{-1}$, but
it steadily increased with time. Due to an occultation of the source
by the Earth, the data are interrupted. However, when the source could
be observed again with {\it RXTE}, it was clear that the count rate
increase had continued. This continuation can still be seen in
beginning of the second data set in Figure~\ref{fig:lc_pca}{\it b}.
The source then reached its highest count rate ($\sim$165 counts
\pers~PCU$^{-1}$) at which approximately level it stayed for the
remainder of the observation.  Similar behaviour was not observed
during the other observations. Hereafter, we refer to the rising part
of this observation as 'the rise' or observation 6a, and to the rest
as 'the plateau' or observation 6b.

\subsection{The rapid X-ray variability \label{section:variability_pca}}

We made power spectra, using 16 s data segments, of the combined
GoodXenon mode data of the AO2 observations and of the AO4 event mode
data in order to study the X-ray fluctuations. No significant rapid
X-ray variability was observed in the individual observations for
frequencies above 0.1 Hz, with typical rms amplitude upper limits of
1\%--3\% on a power-law noise component with index 1 (for the 2--60
keV energy range and a 0.1--10 Hz frequency range). These limits are
consistent with what has been measured by Nowak \& Wilms (1999) using
only the AO2 data.  In order to study the variability at lower
frequencies (a possible very-low frequency noise component or VLFN),
we also made power spectra using 1024 s data segments. Again, only
upper limits could be determined for most observations, except for
observation 6a during which a band-limited noise component was
present. Although the statistics did not allow for constraints on the
exact shape of this noise component, we fitted it with a
power-law. The fit parameters obtained are listed in
Table~\ref{tab:rms}. The noise during this observation was
significantly stronger ($\sim$6\% rms; 0.001--10 Hz) than the upper
limits (95\% confidence levels) obtained for the other observations
(typically a few percent), in particular for observation 6b which
followed observation 6a immediately. However, this strong noise during
observation 6a is very likely caused by the rising trend in this
observation.
\begin{figure}
\begin{center}
\begin{tabular}{c}
\psfig{figure=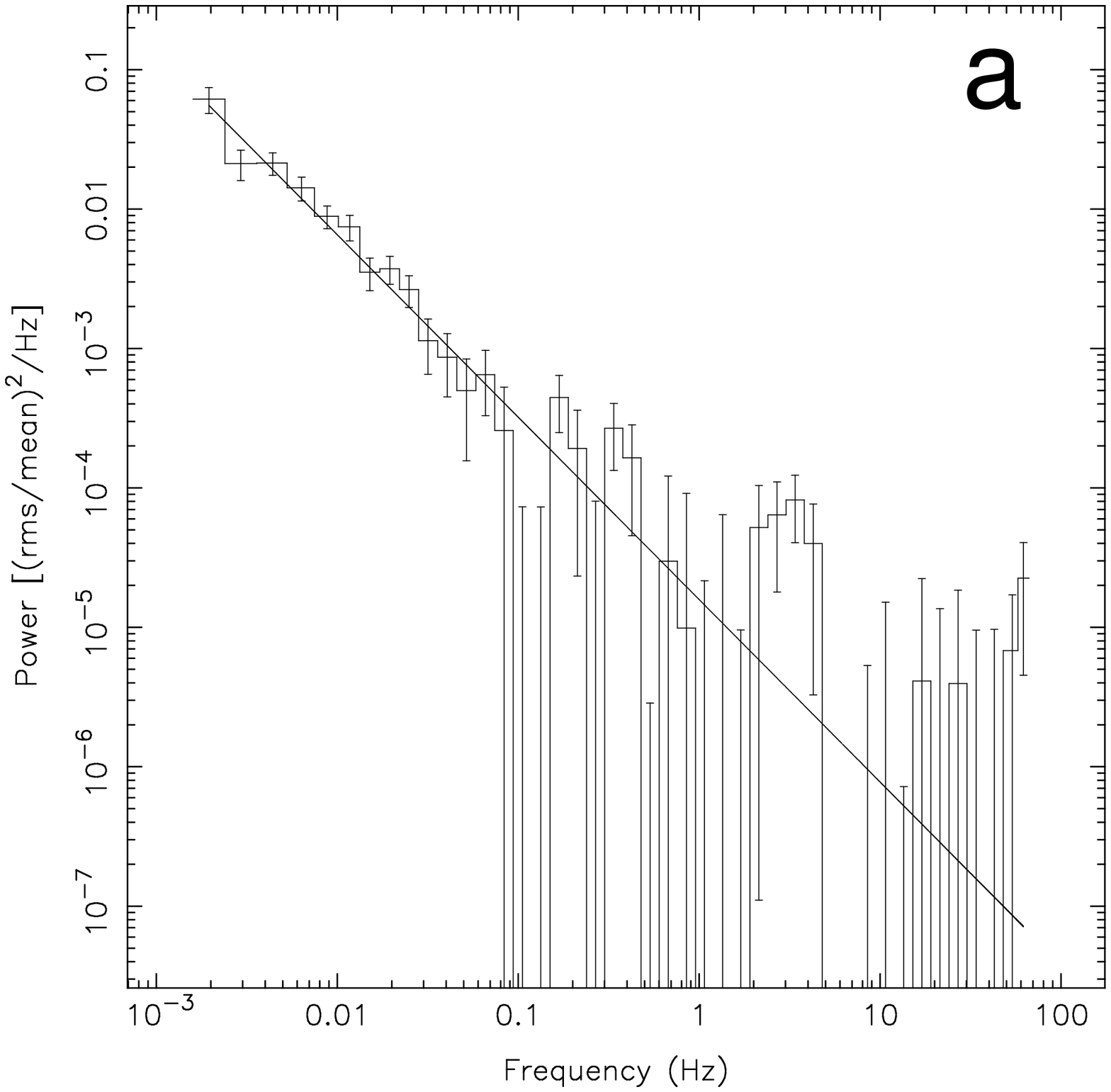,width=4cm}\psfig{figure=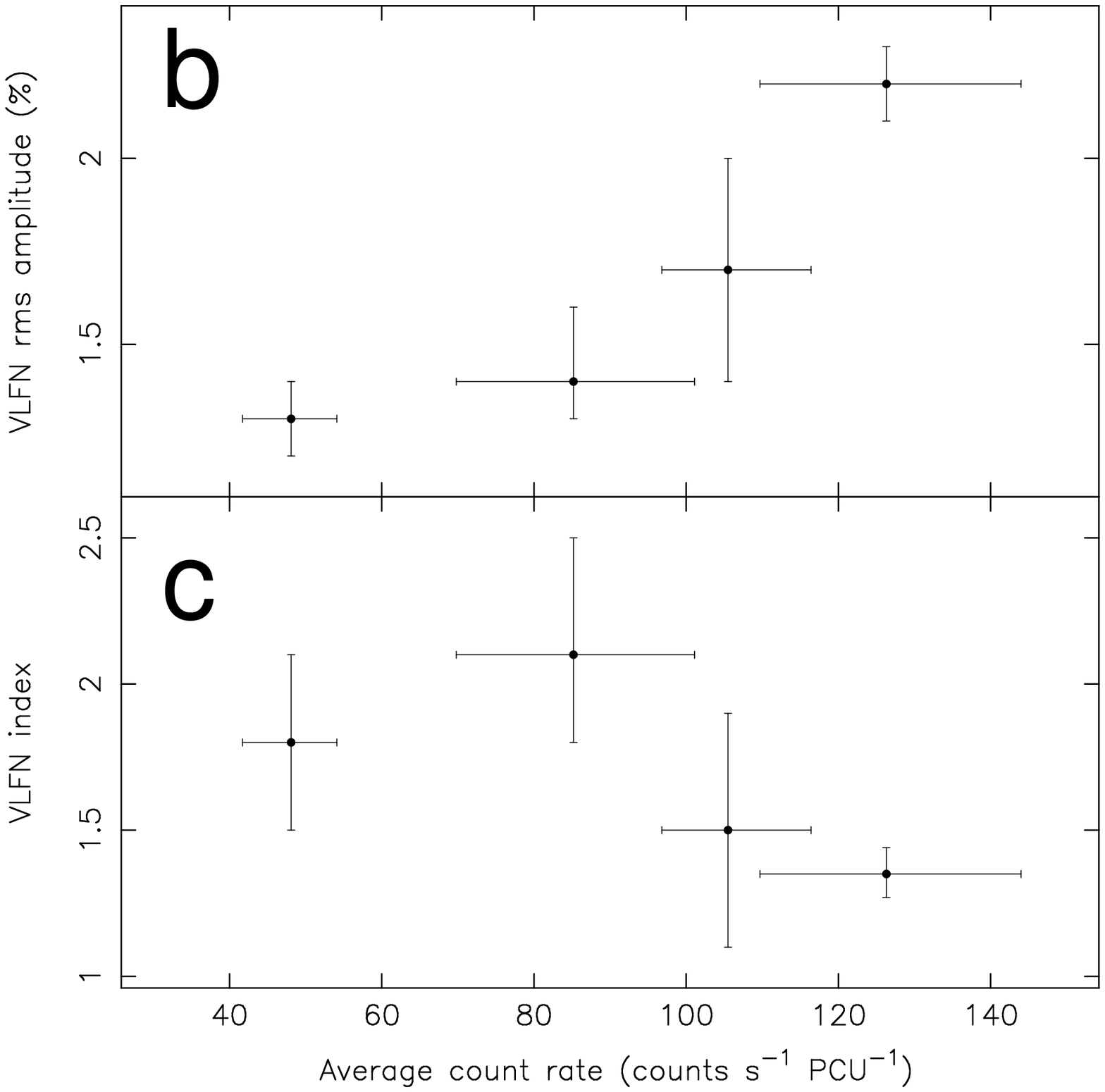,width=4cm}
\end{tabular}
\caption{({\it a}) Power spectrum of all the AO4 observations combined
(except observation 6). The Poisson level has been subtracted. The
VLFN rms amplitude is 1.82\%\pp0.07\% (0.001--10 Hz; 2--60 keV) and
the VLFN index 1.36\pp0.07.  ({\it b}) the VLFN rms amplitude measured
over the frequency range 0.001--10 Hz and ({\it c}) the VLFN index
versus the average 2--60 keV count rate.
\label{fig:power_spectrum} }
\end{center}
\end{figure}

In order to improve our sensitivity for VLFN we combined all the AO4
observations together (except observation 6) to see if a VLFN
component is present below 0.1 Hz. The resulting power spectrum is
shown in Figure~\ref{fig:power_spectrum}{\it a}. We indeed detect a
VLFN with a rms amplitude of 1.82\%\pp0.07\% (0.001--10 Hz; 2--60 keV)
and an index of 1.36\pp0.07. For a source as weak as 4U 1957+11, part
(if not all) of the observed variability might be due to the
fluctuations in the background. To investigate this possibility, we
made similar power spectra of a large number of {\it RXTE}/PCA
background observations performed near the dates of our 4U 1957+11
observations. A power-law shaped noise component was indeed detected
with an rms amplitude of 1\%--2\%. Although this noise is of similar
strength as the VLFN noise component in 4U 1957+11, its index
($\sim$2) is considerably larger than that observed for 4U
1957+11. Furthermore, a large amount of extra source counts are
present in the 4U 1957+11 and if all the variability observed for 4U
1957+11 would have been only due to background fluctuations then the
observed strength of the VLFN component should be weaker than we
observe. Therefore, we conclude that at least part of the observed
variability is due to intrinsic variability of 4U 1957+11.

In order to determine if the strength and the steepness of this noise
component are correlated with the count rate, we combined only those
observations which had approximately similar count rates. We obtained
four count rate intervals and we could detect the VLFN in all of these
selections (see Tab.~\ref{tab:rms}). Note that we did not use
observation 6 in this analysis because of the unusual behaviour of 4U
1957+11 during this observation compared to the other observations.
The rms amplitude and the index of the VLFN are plotted in
Figures~\ref{fig:power_spectrum}{\it b}
and~\ref{fig:power_spectrum}{\it c} versus the average count rate of
the selection.  The range in count rate observed during the
observations which were combined, served as an error on the count
rates. This is a rather conservative error and causes the error bars
on the average count rate to overlap in
Figure~\ref{fig:power_spectrum}{\it b} and
~\ref{fig:power_spectrum}{\it c}. Yet, a clear trend is visible in
that the strength of the VLFN increased with increasing count rate. As
explained above, not all the observed variability of 4U 1957+11 can be
due to background fluctuations. The fact that we observe the strongest
variability when the source is brightest, is also consistent with this
interpretation (if the background was the dominant source of the
variability, it would be expected that the strength would decrease
with count rate as more and more source counts are added). However, at
the lowest count rate selections, it is possible that most of the
variability is due to background fluctuations, but this will only make
the trend in Figure~\ref{fig:power_spectrum}{\it b} more apparent.
The fact that the VLFN index seems to increase when the count rate
decreases also indicates that the variability becomes background
dominated at the lowest count rates.

From Figure~\ref{fig:power_spectrum}{\it b} it is also clear that
observation 6 does not follow the trend. During the rise part of this
observation the average count rate was $\sim$101 counts
\pers~PCU$^{-1}$ but the VLFN strength was $\sim$6\% rms, considerably
stronger than expected from this correlation. We note again that this
strong VLFN component is very likely due to the rising trend.  However
during the plateau phase of this observation the upper limit (95\%
confidence level) on a VLFN component was only 1.1\% rms which was
lower than expected from this correlation. Clearly, during observation
6, 4U 1957+11 behaved quite differently compared to its behaviour
during the other observations.
\begin{figure}
\begin{center}
\begin{tabular}{c}
\psfig{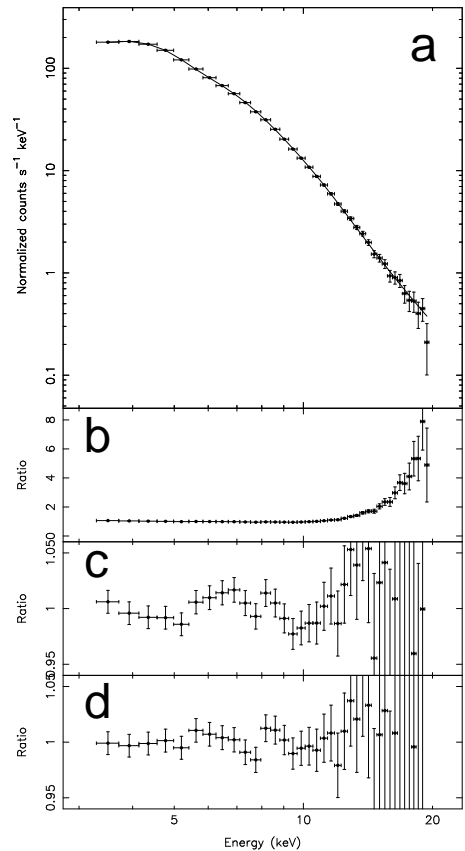}
\end{tabular}
\caption{A typical spectrum of 4U 1957+11 ({\it a}; observation
11). The ratio of the spectrum with a fit consisting of a multi-colour
disc model ({\it b}), a multi-colour disc plus power-law model ({\it
c}), and a multi-colour disc plus power-law model and including a line
at 6.5 keV ({\it d}).
\label{fig:spectrum} }
\end{center}
\end{figure}

\subsection{The X-ray spectra \label{section:spectra_pca}}

We extracted the spectral data obtained in the Standard2f mode to
investigate the spectral characteristics of 4U 1957+11 in each
observation.  Due to the variable number of active detectors and to
obtain the fullest spectral response, we chose to fit the data from
all layers of all active PCUs in each observation.  Background
subtraction and the creation of response matrices were accomplished
using the tools available through LHEASOFT version 5.0 (``pcabackest''
using the ``bright source'' background model, and ``pcarsp,''
respectively).  Fits to an observation of the Crab (nearly a pure
power-law at PCA resolution) made near to the time of the AO4
observations of 4U 1957+11 indicate that the response matrices perform
well above 3 keV, and we therefore adopt this as the lower bound on
our fitting range. Those fits also show energy dependent residuals
which can be as large as 1.0\%; we therefore add 1\% systematic errors
when fitting data from 4U 1957+11 (this has become standard in fitting
PCA data, see, e.g., Miller et al.\ 2001, McClintock et al.\ 2001).
We adopt 20 keV as an upper bound for our fits to 4U 1957+11, as the
spectrum is background-dominated above this energy.  As the {\it
RXTE}/PCA is not very sensitive to the absorption column towards 4U
1957+11, we fixed the $N_{\rm H}$ to $1.3\times10^{21}$ atoms
cm$^{-2}$ (Dickey \& Lockman 1990). Errors on the fit parameters were
calculated for 90\% confidence.
\begin{figure}
\begin{center}
\begin{tabular}{c}
\psfig{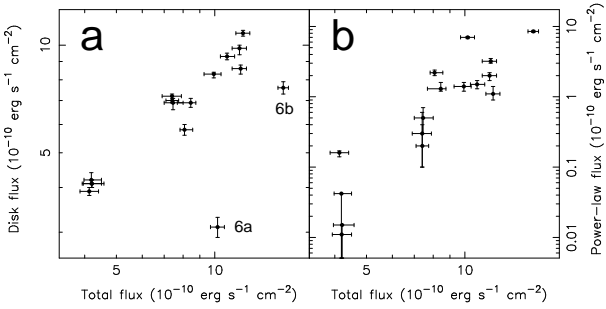}
\end{tabular}
\caption{The disc flux versus the total unabsorbed flux ({\it a}) and
the power-law flux versus the total unabsorbed flux ({\it b}). The
fluxes are for 3--20 keV. The two data points labeled with 6a and 6b
are the data points corresponding to observations 6a and 6b,
respectively (see Tab.~\ref{tab:log}).
\label{fig:fluxes_total} }
\end{center}
\end{figure}

The X-ray spectra of LMXBs and of 4U 1957+11 in particular can be fit
with a variety of models (e.g., Christian \& Swank 1997; Nowak \&
Wilms 1999). However, 4U 1957+11 is a potential BHC and the X-ray
spectra of these sources are usually fitted with a two-component
model: a multi-colour disc black-body (MCD) and a power-law. A similar
consensus is not present for the models for neutron-star
systems. Therefore, we only use the MCD plus power-law model for 4U
1957+11. If this source contains a neutron star, we might obtain
unusual results using this model.

Our fits to the Crab data shows a slight excess near 6 keV with an
equivalent width (EW) of about 40--60 eV. In order to obtain
acceptable fits, we also included a line in our 4U 1957+11 spectral
fits with a fixed energy of 6.5 keV. The obtained EWs of these lines
are considerably larger than 40--60 eV (see Tab.~\ref{tab:spectral}),
which suggest that a line or a complex of lines might be present
around 6.5 keV. Including the line, the fit had 39 degrees of freedom
(41 without line). The reduced $\chi^2$ of the fits were below (but
near) 1 except for observations 1 and 2 were they were $\sim$1.7. A
typical spectrum of 4U 1957+11 is shown in
Figure~\ref{fig:spectrum}{\it a} (see also Nowak \& Wilms 1999 for
another spectrum of 4U 1957+11, but without a power law) and the fit
parameters are listed in Table~\ref{tab:spectral}.

\begin{figure}
\begin{center}
\begin{tabular}{c}
\psfig{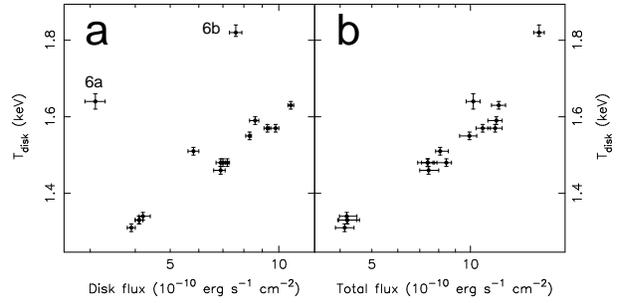}
\end{tabular}
\caption{The disc temperature versus ({\it a}) the disc flux and
({\it b}) the total unabsorbed flux. The fluxes are for 3--20 keV. The
two data points labeled with 6a and 6b are the data points
corresponding to observations 6a and 6b, respectively (see
Tab.~\ref{tab:log}).
\label{fig:tdisc_fluxes} }
\end{center}
\end{figure}

From Figure~\ref{fig:spectrum}{\it b--d} it is clear that a power-law
component was required in most observations (except for the AO2
observations; upper limits on the flux from a possible power-law
component were obtained by fixing the power-law index to 2.5) and that
the addition of a line at 6.5 keV also improved the fit (in this
particular instance the reduced $\chi^2$ decreased from 2.1 without
the line to 0.4 with the line). Some results of the fits are plotted
in Figure~\ref{fig:fluxes_total} and \ref{fig:tdisc_fluxes}. From
Figure~\ref{fig:fluxes_total}{\it a} it is clear that, when not taking
observation 6 into account, the disc flux steadily increases with the
total unabsorbed flux. It is also clearly visible that the data of
observation 6 (both the rise and the plateau part) are significantly
below this correlation. In Figure~\ref{fig:fluxes_total}{\it b}, it
can be seen that the flux in the power-law tends to increase when the
overall flux increases, and the data during observation 6 have the
largest power-law fluxes.  Only in this observation the power-law flux
equals the disc flux, but for the other observations the power-law
flux/disc flux ratio is well below 0.5. Interesting is also the
correlation between the temperature of the disc with the disc flux
(Figure~\ref{fig:tdisc_fluxes}). The temperature is correlated with
the disc flux except again for the data obtained during observation
6. However, when using the total unabsorbed flux instead of the disc
flux, the data of observation 6 line up much better.

\subsection{The X-ray colours \label{section:colours}}

\begin{figure}
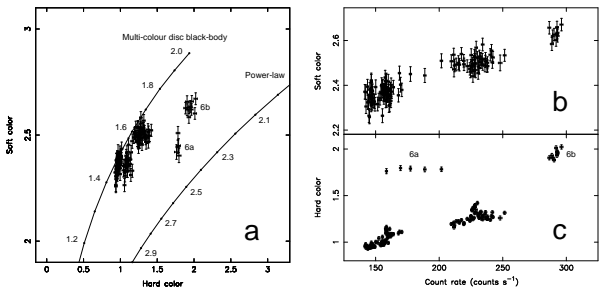

\begin{center}
\begin{tabular}{c}
\psfig{figure=f8a.eps,width=4cm}
\psfig{figure=f8b.eps,width=4cm}
\end{tabular}
\caption{({\it a}) Colour-colour diagram, ({\it b}) soft
hardness-intensity diagram, and ({\it c}) hard hardness-intensity
diagram of 4U 1957+11. The count rate is for 2.9--16.2 keV and 2
detectors. The soft colour is the count rate ratio between 3.7--6.3
keV and 2.9--3.7 keV, and the hard colour the one between 6.3--16.2
keV and 2.9--3.7 keV. The lines in ({\it a}) indicate the colours a
pure multi-colour disc black-body or a pure power-law would have
(assuming a column density of $1.3\times10^{21}$ atoms cm$^{-2}$). In
({\it a}) and ({\it c}), the data points corresponding to observation
6a and 6b are indicated. The count rates used to produce these figures
were background subtracted.
\label{fig:cd_hid} }
\end{center}
\end{figure}

We made a colour-colour diagram (CD; Fig.~\ref{fig:cd_hid}{\it a}) and
hardness-intensity diagrams (HIDs; Figs.~\ref{fig:cd_hid}{\it b} and
{\it c}) using the AO4 Standard2f data (except for observation 5). We
used only the data of the two detectors (PCUs 0 and 2) which were
always active.  The colours used to produce these diagrams are listed
in the caption of Figure~\ref{fig:cd_hid}. Note that both colours have
the same soft band. This was done to ensure the linearity of the CD:
any linear combination of two spectral models lies on the
straight-line segment connecting their locations in the CD (see
Belloni et al. 2000 and Homan et al. 2001 for this procedure). In the
CD, the lines of a pure black-body spectrum and a pure power-law
spectrum are included. The AO2 data and observation 5 were obtain
during gain epoch 3 and the other observations during epoch 4. Due to
the different responses of the PCUs during each epoch, the CDs
obtained for each epoch cannot be compared directly with each
other. We do not show the epoch 3 CD because the AO2 data all fall in
the same place and the long-term gain drift in the PCUs make it
difficult to compare observation 5 with the AO2 data.

From Figure~\ref{fig:cd_hid}{\it a} it is clear that, except for
observation 6, the source spectra are dominated by the MCD component
with only little contribution of the power-law component. The
evolution (i.e., hardening) of the source spectra during observation 6
is mainly due to the increase of the temperature of the MCD component,
but with a significant contribution of an decrease of the power-law
index.  All these features are in good accordance with the results
obtained via the spectral fits (section~\ref{section:spectra_pca},
Tab.~\ref{tab:spectral}). The soft colour and the hard colour (except
observation 6) are also well correlated with the count rate
(Fig.~\ref{fig:cd_hid}{\it b} and {\it c}). The hard colour for
observation 6a is significantly larger than during the other
observations with similar count rates. It is unclear whether the hard
colours of observation 6b are correlated in the same manner with the
count rate as the hard colours of the other observations, or that this
observation by chance lines up with the other observations. From
Fig.~\ref{fig:cd_hid}{\it c} it can also been seen that the hard
colours during the other observations are not always completely
correlated with the count rate; small excursions to higher hard
colours at the same count rates can be observed.

\section{The distance to 4U 1957+11 \label{SECTION:DISTANCE}}

Margon et al. (1978) estimated the distance to 4U 1957+11 to be 7
kpc. They obtained this distance estimate by comparing the optical and
X-ray properties of 4U 1957+11 to that of the bright neutron star LMXB
Sco X-1. Although a valid method at the time, nowadays it is clear
that Sco X-1 and 4U 1957+11 are likely different kinds of objects and
a direct comparison of their properties will likely produce invalid
results.  Moreover, Margon et al. (1978) used an assumed distance to
Sco X-1 of only 500 pc. However, recent high-resolution parallax
measurements of Sco X-1 showed that the most likely distance to this
source is 2.8\pp0.3 kpc (Bradshaw, Fomalont, \& Geldzahler
1999). Following the reasoning of Margon et al. (1978) this revised
distance would result in a new distance estimate of 29 kpc for 4U
1957+11. This distance seems highly unlikely demonstrating that Sco
X-1 and 4U 1957+11 are indeed different types of objects. Therefore,
we regard the distance to 4U 1957+11 as unknown. We note however that
this does not mean that we assume that 4U 1957+11 is further away than
7 kpc and thus more luminous than previous thought. In fact the source
might be much closer and might be less luminous than $10^{36-37}$
\Lunit.

\section{Discussion\label{discussion}}

We have observed 4U 1957+11 at a variety of X-ray luminosities (using
both the {\it RXTE}/ASM and PCA data) and studied its spectral and
temporal behaviour at those different luminosities. We first compare
our results with those obtained previously for 4U 1957+11 and then
with other LMXBs.

\subsection{Comparison with other 4U 1957+11 observations}

Our spectral results for 4U 1957+11 agree very well with those listed
in the literature for this source (e.g., Yaqoob et al. 1993; Ricci et
al. 1995; Nowak \& Wilms 1999). In particular, the same hardening of
the source spectrum we see with increasing flux had already been found
earlier (e.g., Yaqoob et al. 1993; Ricci et al. 1995). Yaqoob et
al. (1993) even found the same small excursion of the source in the
HID towards harder colours at the same intensity as we do (e.g.,
compare their Figure 2 with our Figure~\ref{fig:cd_hid}; here we do
not mean the enigmatic behaviour of observation 6, but the very small
excursions in the HID mentioned at the end of
section~\ref{section:colours}).  The lack of rapid X-ray variability
reported in the literature (e.g., Ricci et al. 1995; Nowak \& Wilms
1999) is also consistent with the very low amplitude noise we found.
The only unusual observation in our data is observation 6 for which no
similar observation has been reported in the literature. Besides this
observation, the behaviour of 4U 1957+11 seems to be very stable over
many years as observed with several different X-ray instruments.

Nowak \& Wilms (1999) claimed a 177 day period in 4U 1957+11. As shown
above (\S~\ref{section:asm}), the long-term X-ray variability of this
source is much more complicated than one single period. Also, we found
that the luminosity of 4U 1957+11 can change by a factor of at least
4, which is significantly larger than what had been observed
previously for this source.  These variations cannot be due to changes
in the column density because the high column densities ($\sim10^{23}$
cm$^{-2}$) needed to account for such large luminosity fluctuations
would easily have been observed in our spectra.  These findings
suggest that the reason behind the long-term variability is most
likely not a precessing accretion disc as suggested by Nowak \& Wilms
(1999) but might have another cause, possibly variations in the mass
accretion rate onto the compact object.  This conclusion seems to
confirm a recent theoretical study about precessing warped accretion
discs in X-ray binaries.  Ogilvie \& Dubus (2001) showed that the
orbital radius of 4U 1957+11 is well below the critical radius for a
stable precessing accretion disc to evolve in this system. A possible
mechanism for the long-term fluctuations in systems like 4U 1957+11
might be that a modulation of the accretion rate through the disc can
occur when a viscously unstable disc is irradiated by a constant X-ray
flux (Ogilvie \& Dubus 2001; G. Dubus 2001 private communication; see
also Dubus et al. 2001 in preparation).

\subsection{Comparison with neutron-star LMXBs}

The behaviour displayed by 4U 1957+11 is not that typical for
neutron-star LMXBs.  For such neutron-star systems, the X-ray spectrum
can also be described by a soft component below 3 keV and a power-law
component above 10 keV. However, the power-law component becomes
significantly weaker when the luminosities or the inferred mass
accretion rates of these systems increase (see, e.g., Di Salvo et
al. 2000 and references therein), which is completely opposite to what
we observe for 4U 1957+11.  It is possible that for 4U 1957+11 the
mass accretion rate is anti-correlated with the X-ray luminosity as
observed in the brightest neutron star LMXBs (the so-called Z sources;
see Hasinger \& van der Klis 1989), but we regard this possibility as
unlikely. At times when the luminosity is thought to be
anti-correlated with the mass accretion rate in Z sources (on their
``normal branch''), the intrinsic luminosity is close to the Eddington
luminosity, but 4U 1957+11 seems to be significantly weaker (the
source has to be unlikely distant in order to have near-Eddington
luminosities).

Also, quasi-periodic oscillations near 7 Hz and 50--60 Hz are often
reported in the Z sources when they are on their normal branches
together with band-limited noise which increases in strength when the
source becomes harder (up to 10\%--15\% rms amplitude; see van der
Klis 1995 for a review about the timing properties of Z sources). Such
strong noise and quasi-periodic oscillations have never been observed
for 4U 1957+11, despite the range in luminosities sampled and the
different instruments used (Ricci et al. 1995; Nowak \& Wilms
1999). We tentatively suggest that it is more likely that 4U 1957+11
harbors a black hole than a neutron star.

\subsection{Comparison with black-hole X-ray binaries \label{section:bhcs}}

It has already been noted that the soft X-ray spectrum of 4U 1957+11
resembles that of BHCs when they are in their high state (e.g., White
\& Marshall 1984; see Homan et al. 2001 for a recent detailed
discussion of black-hole states). The very low amplitude variability
observed for 4U 1957+11 is also consistent with this. Several
arguments were put forward against a black-hole nature of 4U
1957+11. A main argument is that the luminosity of 4U 1957+11 is
considerably lower than that of BHCs in their high state. But, as
already explained above (\S~\ref{SECTION:DISTANCE}), the distance to
the source is not well known. Although the true luminosity of 4U
1957+11 could be considerably higher than the normal 5--10
$\times10^{36}$ \Lunit~assumed, it could also be less. However, for
several BHCs (e.g., GRO J1655--40; XTE J1550--564; M\'endez, Belloni,
\& van der Klis 1998, Sobczak et al. 1999; Homan et al. 2001) it has
been found that high state episodes can occur also in those sources at
relatively low luminosities ($<10^{37}$ \Lunit). This demonstrates
that the luminosity argument against 4U 1957+11 being a black hole is
not well founded.

Another argument is that the obtained inner disc radius obtained from
the spectral fits (several km) is too low and the temperature of the
disc ($\sim1.5$ keV) too high to be in agreement with what has been
found for other BHCs (e.g., Yaqoob et al. 1993). However, the disc
radius depends on the unknown distance and inclination of the system
and they might be larger than usually assumed (see also Nowak \& Wilms
1999). Disc temperatures close to those observed for 4U 1957+11 have
also been observed during the high states of other BHC, such as LMC
X-3 (see below) and XTE J1748--288 (Miller et al. 2001). For the
enigmatic BHC GRS 1915+105 disc temperatures up to 2 keV have been
observed during its high state (e.g., Rao, Yadav, \& Paul 2000),
although it is unclear if a comparison between 4U 1957+11 and GRS
1915+105 is really valid because of the very unusual behaviour of GRS
1915+105. However, this demonstrates that although the disc
temperature of 4U 1957+11 is relatively high compared to most BHCs, it
is not a compelling reason to dismiss the possibility that 4U 1957+11
contains a black hole. Especially a comparison of this source with the
strong black-hole X-ray binary LMC X-3 suggests a black-hole nature
for 4U 1957+11.

\subsection{Comparison with LMC X-3 \label{section:lmcx-3}}

LMC X-3 is a strong BHC with a black-hole mass of $>$5.8 \mzon~(see
Soria et al. 2001 and references therein). This source is unusual in
that it spends most of its time in the canonical black-hole high state
although occasionally it makes excursions to the black-hole low state
(see Boyd et al. 2000, Wu et al. 2001, and Wilms et al. 2001 for
observations of LMC X-3 in its low state).  The spectrum of LMC X-3 is
very similar to that of 4U 1957+11 (although its disc temperature is
at maximum only $\sim$1.3 keV; Nowak et al. 2001; Wilms et
al. 2001). Its spectrum hardens with increasing luminosity (Cowley et
al. 1991; Ebisawa et al. 1993; Wilms et al. 2001; note that dramatic
spectral hardening also occur when the source luminosity decreases
considerably, thus when the source transit from its normal high state
to its low state; see Boyd et al. 2000 and Wilms et al. 2001) in a way
similar to what we observed for 4U 1957+11 (the hardening is mostly
due to an increase of the disc temperature). And as for 4U 1957+11,
hardly any variability on short time scale could be detected (e.g.,
Nowak et al. 2001) and the source exhibits strong long-term X-ray
fluctuations (Cowley et al. 1994; Wilms et al. 2001; factor of 5 for
LMC X-3 and at least 4 for 4U 1957+11). The only apparent differences
between the two sources is the fact that LMC X-3 can be as bright as
$10^{38}$ erg \pers~(2--10 keV) and that 4U 1957+11 is most likely
considerably weaker. As explained above (\S~\ref{SECTION:DISTANCE}),
the distance to 4U 1957+11 is not well-constrained and the source
could be further away than the (usually) assumed distance of 7 kpc
(Margon et al. 1978). However, this last effect might only account for
a small increase in the luminosity of 4U 1957+11 because the source
should be far away to have an intrinsic luminosity similar to what is
usually observed from LMC X-3.  

It is possible that the black-hole mass in 4U 1957+11 is considerably
lower than that of the black hole in LMC X-3. If the specific
accretion rate above which a black-hole X-ray binary enters the high
state is tied to the mass of the black hole (e.g., such transition
occurs at a fixed fraction of the critical Eddington mass accretion
rate), then it will be reached earlier in 4U 1957+11 than in LMC
X-3. This could explain why 4U 1957+11 is not as bright as LMC X-3.  A
possible other solution might be the lower metallicity in the LMC
compared to the Galaxy. This might give higher luminosities for
accreting X-ray binaries (as proposed to explain the on averaged
higher luminosities for the X-ray binaries in the SMC and LMC; see,
e.g., Clark et al. 1978). It remains to be determined whether this
could explain the luminosity difference between the two sources.

Despite this difference in intrinsic luminosity, the many similarities
between these two sources suggest that, similar to LMC X-3, 4U 1957+11
harbors a black hole and the long-term fluctuations are due to changes
in the mass accretion rate. If this can be proven, then 4U 1957+11
would be the only persistent Galactic BHC which is continuously
observed in the black-hole high state. The other persistent galactic
BHCs (e.g., Cyg X-1, 1E 1740.7--2942) are most often observed in the
black-hole low state and only rarely make excursions to the high state
(the other persistent BHC which spends all of its time in the
black-hole high state is LMC X-1, also in the LMC, and is a wind-fed
system).

Several predictions can be made if we assume that 4U 1957+11 is indeed
is black-hole binary. It is possible that at very high luminosities
(higher than so far observed), the source might transit to the
black-hole very high state and strong quasi-periodic oscillations
might be observed. At the lowest end of the luminosity range, it might
be possible that the source will enter the black-hole low state and
the spectrum will then considerably harden and strong rapid X-ray
variability will be observed. Such a transition was recently observed
for LMC X-3 (Boyd et al. 2000; Wilms et al. 2001) suggesting that such
a change is also possible for 4U 1957+11. Observing either state in 4U
1957+11 would strongly confirm that this source is a BHC.

It is possible that the excursion of the source to a harder state
during observation 6, can be identified with a very high state, but
the weak variability of the source during this observation and the
non-detections of quasi-periodic oscillations around 6 Hz does not
seem to be consistent with this scenario.  However, in other BHCs
strong variability or quasi-periodic oscillations have not always been
observed when they were at high luminosities and exhibited hard
spectra. For example, for XTE J1550--564 it was observed that, when
this source was relatively bright, it occasionally exhibited short
excursions to hard states without a significant increase in the
strength of the rapid X-ray variability (see Homan et al. 2001 and in
particular the small flares discussed by these authors). Also,
Wijnands et al. (2001) reported that only weak noise (less than a few
percent rms amplitude) could be detected during some very-high state
observations of GRS 1739--278. Therefore, it is still possible that 4U
1957+11 might have exhibited very-high state like behaviour during
observation 6.

If indeed 4U 1957+11 exhibits state behaviour similar to other BHCS,
then during those hard states an increase in radio luminosity is also
expected and a monitoring radio campaign on the source might be able
to reveal radio emission from this source at the extreme end of the
luminosity range. The only radio observations of 4U 1957+11 reported
so far, did not reveal any strong radio emission from the source
(Nelson \& Spencer 1988), however, it is unclear what the X-ray
luminosity was at the time of those observations. However, our {\it
RXTE} observations show that 4U 1957+11 spends most (if not all) of
its time in the soft state and it is likely it was in a similar state
during the Nelson \& Spencer (1988) observation. This would be
consistent with their non-detection in the radio band, because
black-hole candidates do not show strong radio emission in their soft
states (e.g., Fender 2001).  Note also, that null results from a
monitoring radio campaign will not exclude the possibility that 4U
1957+11 is a BHC because it might well be possible that the high or
low accretion rates needed for the state changes to occur are not
reached in this system.

\section{Conclusion}

The conclusions obtained from our analysis of the {\it RXTE} data of
4U 1957+11 are:

\begin{itemize}
        \item{The X-ray luminosity of 4U 1957+11 can vary by at least
        a factor of 4.}

	\item{The 3--20 keV spectrum can be adequately described by a
	multi-colour disc black-body and a power-law component.}

	\item{The spectrum significantly hardens when the luminosity
	increases, both as a result of an increase of the disc
	temperature and, although to a lesser degree, an increase in
	the power-law flux contribution to the total flux.}

       \item{Only very weak variability on short time scales could be
	observed (of order one percent rms amplitude).}

	\item{A comparison of 4U 1957+11 with other X-ray binaries,
	i.e., LMC X-3, suggests that 4U 1957+11 contains a black hole
	and not a neutron star. The spectral and temporal properties
	of 4U 1957+11 indicate that the source is a black-hole LMXB
	persistently in the black-hole high state. If confirmed, the
	source would be the only such source known.}

	\item{The long-term variability of the source is highly
	complex and most likely due to variations in the accretion
	rate onto the black hole.}

\end{itemize}

\section*{Acknowledgments}

This work was supported by NASA through Chandra Postdoctoral
Fellowship grant number PF9-10010 awarded by CXC, which is operated by
SAO for NASA under contract NAS8-39073. This work was also supported
in part by the Netherlands Organization for Scientific research (NWO).
This research has made use of data obtained through the HEASARC Online
Service, provided by the NASA/GSFC and quick-look results provided by
the ASM/{\it RXTE} team.  RW thanks Guillaume Dubus for useful
discussions.

\begin{table*}
\caption{Log of the observations \label{tab:log}}
\begin{flushleft}
\begin{tabular}{llccccc}
\hline
\hline
Number &Observation ID & Observation time & Good time & PCUs on & Count rate$^a$               & Count rate range$^b$       \\
                  &               &      (UTC)       & (ksec)    &         & (counts s$^{-1}$ PCU$^{-1}$) & (counts s$^{-1}$ PCU$^{-1}$)\\
\hline
\multicolumn{7}{c}{AO2 observations}\\
   & 20184-$^c$  &                          & 29.2    &     & 48.1\pp0.05            & 41.7--54.1 \\
1  & ~~01-01-000 & 1997 Nov 26 00:23--07:45 & 15.5    & All & 48.4\pp0.1             & 42.7--54.1 \\
2  & ~~01-01-00  & 1997 Nov 26 08:23--10:44 & 6.2     & All & 48.1\pp0.1             & 43.0--52.7 \\
3  & ~~01-01-01  & 1997 Nov 27 01:59--04:05 & 4.7     & All & 47.4\pp0.1             & 41.7--53.1 \\
4  & ~~01-01-02  & 1997 Nov 29 02:00--02:48 & 2.8$^d$ & All/1,2,3,4$^d$ & 48.2\pp0.1 & 43.9--53.4 \\
\hline
\multicolumn{7}{c}{AO4 observations}\\
   & 40044-$^c$ &  & 38.6 & & 105.6\pp0.05 & 69.8--171.   \\
5  & ~~01-01-00 & 1999 March 03 02:33--03:19  & 2.7 & All & 105.6\pp0.1              & 96.9--116.5 \\ 
6  & ~~01-02-01$^e$ & 1999 April 03 10:43--12:42  & 3.9 & All & &  \\
~6a& ~~~~Rise   & 1999 April 03 10:43--11:06  & 1.3 &     & 101.1\pp0.2              & 83.7--123.6\\
~6b& ~~~~Plateau& 1999 April 03 11:43--12:42  & 2.6 &     & 160.9\pp0.2              & 150.8--171.4 \\
7  & ~~01-02-00 & 1999 April 06 23:01--23:50  & 3.0 & 0,2,3 & 131.2\pp0.2            & 118.4--144.9 \\
8  & ~~01-03-02 & 1999 April 17 03:41--04:28  & 2.8 & 0,1,2 &  86.7\pp0.1            & 89.2--93.8  \\
9  & ~~01-03-03 & 1999 April 17 05:18--05:54  & 2.2 & 0,2  & 86.7\pp0.2              & 76.1--94.7   \\
10 & ~~01-03-00 & 1999 April 20 21:03--21:45  & 2.5 & 0,2,4 & 86.5\pp0.2             & 76.5--95.6   \\
11 & ~~01-03-01 & 1999 April 21--22 23:25--01:11 & 3.9 & 0,2,3,4 & 90.3\pp0.2        & 83.1--101.1  \\
12 & ~~01-04-00 & 1999 May   04 18:12--21:45  & 8.1 & 0,2,3,4 & 81.5\pp0.1           & 69.8--92.0  \\
13 & ~~02-01-02 & 1999 Dec   01 23:05--23:42  & 2.2 & All & 126.0\pp0.2              & 117.5--136.0 \\
14 & ~~02-01-01 & 1999 Dec   02 00:39--01:24  & 2.6 & All & 128.8\pp0.2              & 118.6--144.0 \\
15 & ~~02-01-00 & 1999 Dec   02 02:15--04:19  & 4.7 & All & 121.9\pp0.2              & 109.7--138.9 \\
\hline
\hline
\multicolumn{7}{l}{$^a$ Averaged, background subtracted count rate per
ObsID  for the nominal {\it
RXTE}/PCA energy range of 2--60 keV}\\
\multicolumn{7}{l}{\,\,\, (all 129 energy channels for the
Standard 2 mode data). The error bars on the count rates are
statistical errors.}\\
\multicolumn{7}{l}{$^b$ The range of count rates observed for each observation.}\\
\multicolumn{7}{l}{$^c$ All observations with this ObsID combined}\\
\multicolumn{7}{l}{$^d$ 1.5 ksec with all PCUs on and 1.3 ksec with only
PCUs 1--4 on}\\
\multicolumn{7}{l}{$^e$ Significant increase in the count rate observed during the observation}
\end{tabular}
\end{flushleft}
\end{table*}

\begin{table*}
\caption{VLFN parameters versus count rate \label{tab:rms}}
\begin{flushleft}
\begin{tabular}{cccc}
\hline
\hline
Average count rate$^a$ & Amplitude$^b$ & Index & Observations\\
         (counts  s$^{-1}$ PCU$^{-1}$) & (\% rms) & & combined \\
\hline
101$^{+23}_{-17}$       &  6.0$^{+0.7}_{-0.6}$ & 0.7\pp0.1               & 6a     \\
160\pp11                &  $<$1.1              & 1.5$^c$                 & 6b      \\
\hline
48\pp6                  &  1.3\pp0.1           & 1.8\pp0.3               & 1--4   \\
85\pp16                 &  1.4$^{+0.2}_{-0.1}$ & 2.1$^{+0.4}_{-0.3}$     & 8--12  \\
106$^{+11}_{-9}$        &  1.7\pp0.3           & 1.5\pp0.4               & 5      \\
126\pp18                &  2.2\pp0.1           &
1.35\pp0.09             & 7, 13--15 \\
\hline
\hline
\multicolumn{4}{l}{$^a$ For the
total 2--60 keV {\it RXTE}/PCA energy range. The errors are the
range}\\
\multicolumn{4}{l}{\,\,\, of count rates observed during 
the observations which were combined. }\\
\multicolumn{4}{l}{\,\,\, The count rates are background subtracted. }\\
\multicolumn{4}{l}{$^b$ Measured for the frequency range 0.001--10 Hz}\\
\multicolumn{4}{l}{$^c$ Index fixed}
\end{tabular}
\end{flushleft}
\end{table*}

\hoffset -1cm

\begin{table*}
\caption{X-ray spectral parameters$^a$ \label{tab:spectral}}
\begin{flushleft}
\begin{tabular}{cccccccccccc}
\hline
\hline
Number &  T$_{disc}$ & Norm$_{disc}$         & F$_{disc}$$^b$        & $\Gamma$                    & Norm$_{pl}$             & F$_{pl}$$^b$          & Width                 & Norm$_{line}$          & EW                    & F$_{line}$$^b$        &  F$_{tot}^c$  \\
       &   (keV)     &                       &                       &                             &    ($\times10^{-2}$)    &                       & (keV)                 & ($\times10^{-4}$)      & (eV)                  &                       &          \\
\hline
1      & 1.33\pp0.01 &  15.6\pp0.30          &  4.1\pp0.1            &   2.5$^d$                   & $<$0.13                 &  $<$0.015             &  0.8\pp0.2            &   4.1\pp0.7            &  107\pp18             &  4.3\pp0.8            &  4.2$^{+0.4}_{-0.3}$ \\
2      & 1.31\pp0.01 &  15.9$^{+0.4}_{-0.3}$ &  3.9\pp0.1            &   2.5$^d$                   & 1.4$^{+0.1}_{-0.2}$     &  0.16$^{+0.01}_{-0.02}$ &  1.0\pp0.2          &   5.2\pp0.9            &  136\pp22             &	 5.4\pp0.9            &  4.1\pp0.3 \\
3      & 1.33\pp0.01 &  15.6\pp0.4           &  4.1\pp0.1            &   2.5$^d$                   & $<$0.10                 &  $<$0.011             &  0.6$^{+0.2}_{-0.3}$  &   3.7\pp0.8            &   98\pp21             &	 3.8\pp0.9            &  4.2\pp0.3 \\
4      & 1.34\pp0.01 &  15.0$^{+0.2}_{-0.1}$ &  4.2$^{+0.2}_{-0.1}$  &   2.5$^d$                   & $<$0.37                 &  $<$0.042             &  0.8$^{+0.7}_{-0.8}$  &   2.3$^{+4.4}_{-2.0}$  &   57$^{+112}_{-49}$   &	 2.5$^{+4.5}_{-2.2}$  &  4.2$^{+0.3}_{-0.2}$ \\
\hline
5      & 1.55\pp0.01 &  14.4\pp0.3           &  8.3$^{+0.1}_{-0.2}$  &   2.32$^{+0.05}_{-0.06}$    &   8.9\pp1.2             &  1.4\pp0.2            &  0.9\pp0.5            &   6.4\pp2.2            &  65\pp22              &   6.5\pp2.1           &	 9.9$^{+0.5}_{-0.7}$ \\
6a     & 1.64\pp0.02 &   4.0\pp0.3           &  3.1\pp0.2            &   2.42$^{+0.01}_{-0.02}$    &  52.4\pp1.7             & 7.0$^{+0.2}_{-0.3}$   &  0.4$^{+0.5}_{-0.4}$  &   4.8\pp2.1            &  53\pp24              &   5.0\pp2.2           &  10.2\pp0.5           \\
6b     & 1.82$^{+0.02}_{-0.01}$&  5.8\pp0.2  &  7.6\pp0.3            &   2.34$^{+0.01}_{-0.02}$    &   54.9\pp1.8            & 8.5\pp0.3             &  0.3$^{+0.6}_{-0.3}$  &   5.8$^{+3.1}_{-2.7}$  &  40\pp20              &  6.0$^{+3.2}_{-2.8}$  &	16.2\pp0.6          \\               
7      & 1.63\pp0.01 &  14.6\pp0.3           & 10.8\pp0.2            &   2.3\pp0.1                 &   6.8$^{+1.7}_{-1.4}$   &  1.1$^{+0.3}_{-0.2}$  &  0.8\pp0.3            &   9.9\pp2.7            &  79\pp22              &  10.3\pp2.8           &	12.2\pp0.6  \\
8      & 1.48\pp0.01 &  15.3$^{+0.4}_{-0.5}$ &  7.2$^{+0.1}_{-0.3}$  &   2.3\pp0.4                 &   1.6$^{+1.9}_{-0.9}$   &  0.3$^{+0.3}_{-0.2}$  &  0.5$^{+1.0}_{-0.5}$  &   3.6$^{+2.2}_{-1.7}$  &  48$^{+29}_{-23}$     &   3.7$^{+2.3}_{-1.7}$ &	 7.4\pp0.5 \\
9      & 1.46\pp0.01 &  15.8\pp0.5           &  6.9$^{+0.2}_{-0.3}$  &   2.7\pp0.2                 &   6.6$^{+3.2}_{-2.3}$   &  0.5\pp0.2            &  1.4\pp0.3            &  11.6$^{+2.0}_{-2.4}$  & 158$^{+29}_{-33}$     &  12.0$^{+2.1}_{-2.5}$ &	 7.5$^{+0.6}_{-0.5}$ \\
10     & 1.51\pp0.01 &  11.3\pp0.4           &  5.8\pp0.2            &   2.94\pp0.04               &  44.4$^{+4.3}_{-3.6}$   &  2.2\pp0.2            &  0.3$^{+0.4}_{-0.3}$  &   3.7\pp1.7            &  48\pp22              &   3.8\pp1.7           &	 8.1$^{+0.5}_{-0.3}$ \\
11     & 1.48\pp0.01 &  15.1\pp0.4           &  6.9\pp0.2            &   2.48\pp0.06               &  11.5$^{+1.7}_{-1.4}$   &  1.3$^{+0.3}_{-0.1}$  &  0.9\pp0.3            &   8.1\pp1.8            & 100\pp22              &   8.4\pp1.9           &	 8.4$^{+0.3}_{-0.8}$ \\
12     & 1.48\pp0.01 &  15.4\pp0.1           &  7.0\pp0.1            &   2.4\pp0.3                 &   1.5$^{+1.4}_{-0.7}$   &  0.2$^{+0.2}_{-0.1}$  &  0.6\pp0.3            &   4.7\pp1.4            &  64\pp19              &   4.9\pp1.5           &	 7.4\pp0.3 \\
13     & 1.59\pp0.01 &  13.0$^{+0.3}_{-0.4}$ &  8.6$^{+0.2}_{-0.3}$  &   2.41$^{+0.04}_{-0.03}$    &  24.1\pp2.3             &  3.2$^{+0.3}_{-0.2}$  &  0.8\pp0.5            &   7.0\pp2.6            &  60\pp22              &   7.3\pp2.7           &	 12.0$^{+0.5}_{-0.7}$ \\
14     & 1.57\pp0.01 &  15.7$^{+0.3}_{-0.4}$ &  9.8$^{+0.2}_{-0.4}$  &   2.17\pp0.05               &   9.2\pp1.2             &  2.0$^{+0.2}_{-0.3}$  &  1.0\pp0.3            &  10.9\pp2.7            &  91\pp23              &  11.3\pp2.8           &	 11.9\pp0.6 \\
15     & 1.57\pp0.01 &  14.8\pp0.3           &  9.3\pp0.2            &   2.75\pp0.06               &  21.3\pp3.0             &  1.5\pp0.2            &  0.9\pp0.2            &  13.9\pp2.3            & 126\pp22              &  14.4\pp2.4           &	 10.9$^{+0.6}_{-0.5}$ \\
\hline
\hline

\multicolumn{12}{l}{$^a$ The column density was fixed to
$1.3\times10^{21}$ atoms cm$^{-2}$, the errors are for 90\% confidence
levels, the centroid energy of the line was fixed }\\
\multicolumn{12}{l}{\,\,\, at 6.5 keV, and the reduced
$\chi^2$ of the fits were below 1 except for observations 1 and 2 where
they were $\sim$1.7.}\\
\multicolumn{12}{l}{$^b$ Fluxes are for 3--20 keV, absorbed, and in
units of 10$^{-10}$ \funit~ for the MCD and power law component and in}\\
\multicolumn{12}{l}{\,\,\, 10$^{-12}$ \funit~ for the line}\\
\multicolumn{12}{l}{$^c$ Total fluxes are for 3--20 keV, unabsorbed, and in units of 10$^{-10}$ \funit}\\
\multicolumn{12}{l}{$^d$ Parameter fixed}
\end{tabular}
\end{flushleft}
\end{table*}

\end{document}